\documentclass[fleqn,usenatbib,useAMS]{mnras}

\usepackage{graphicx}	% Including figure files
\usepackage{amsmath}	% Advanced maths commands
\usepackage{amssymb}	% Extra maths symbols
\usepackage{multicol}   % Multi-column entries in tables
\usepackage{bm}		% Bold maths symbols, including upright Greek
\usepackage{pdflscape}	% Landscape pages

\usepackage[T1]{fontenc}
\usepackage{ae,aecompl}

\usepackage{newtxtext,newtxmath}
\title{Four newly discovered HII galaxies}

\author[O. Garde et al.]{
O. Garde,$^{1}$\thanks{E-mail: o.garde@free.fr}
P. Le D\^{u},$^{2}$\thanks{E-mail: pascal.e.du@shom.fr}
M. Koenig,$^{3}$\thanks{E-mail: michael@schmid-koenig.de}
P. Dubreuil,$^{4}$
A. Lopez,$^{5}$
B. Guegan$^{6}$
\\
$^{1}$Observatoire de la Tourbi$\grave{e}$re, F-38690 Chabon, France\\
$^{2}$Observatoire de Kermerrien, F-29840 Porspoder, France\\
$^{3}$IAU observatory C87, D-64668 Rimbach, Germany\\
$^{4}$F-06790 Aspremont, France\\
$^{5}$F-06110 Le Cannet, France\\
$^{6}$CALA, F-69120 Vaulx-en-Velin, France
}

\date{Last updated 2015 May 22; in original form 2013 September 5}

\pubyear{2019}

\begin{document}
\label{firstpage}
\pagerange{\pageref{firstpage}--\pageref{lastpage}}
\maketitle

% Abstract of the paper
\begin{abstract}
We present the results of spectroscopy campaigns for planetary nebula candidates, where we have identified four objects as Seyfert galaxies. All observations have been carried out by a group of French amateur astronomers. During the campaigns at the C\^{o}te d'Azur observatory at Calern (France), four HII galaxies could be identified. Using the naming convention of our campaign, these objects are (1) App~1 (RA: 22h 49m 20.23s, DEC: +46\degr 07\arcmin 37.17\arcsec), (2) Pre~21 (RA: 18h 04m 19.62s, DEC: +00\degr 08\arcmin 04.96\arcsec), (3) Pre~24 (RA: 04h 25m 53.63s, DEC: +39\degr 49\arcmin 19.69\arcsec), and (4) Ra~69 (RA: 19h 30m 23.64s, DEC: +37\degr 37\arcmin 06.58\arcsec).
\end{abstract}

% Select between one and six entries from the list of approved keywords.
% Don't make up new ones.

%%%%%%%%%%%%%%%%%%%%%%%%%%%%%%%%%%%%%%%%%%%%%%%%%%

%%%%%%%%%%%%%%%%% BODY OF PAPER %%%%%%%%%%%%%%%%%%

% The MNRAS class isn't designed to include a table of contents, but for this document one is useful.
% I therefore have to do some kludging to make it work without masses of blank space.
%\begingroup
%\let\clearpage\relax
%\tableofcontents
%\endgroup
%\newpage

%"App1" 22 49 20.23 +46 07 37.17
%"Pre21" 18 04 19.62 +00 08 04.96
%"Pre24" 04 25 53.63 +39 49 19.69
%"Ra69" 19 30 23.64 +37 37 06.58

\section{Introduction}
Generally, planetary nebula candidates are discovered by amateur astronomers in two different ways: Either with small diameter amateur instruments with low focal f/d ratio and narrow bands filters H$\alpha$, [OIII] and [SII]. Or, in professional surveys like SDSS\footnote{Sloan Digital Sky Survey: \url{http://www.sdss.org}}, WISE\footnote{Wide-field Infrared Survey Explorer: \url{https://www.jpl.nasa.gov/wise}}, 
sky-map.org\footnote{Interactive Sky-Map: \url{http://www.sky-map.org}}, DECaPS\footnote{Interactive viewer of the DECam Legacy Survey: \url{http://decaps.skymaps.info/viewer.html}}, PanSTARRS\footnote{ Panoramic Survey Telescope and Rapid Response System: \url{https://panstarrs.stsci.edu/}} are used to provide list of candidates, which are then studied in detail in successive spectroscopic campaigns.

We, a group of French amateur astronomers, have compiled a list of planetary nebula candidates. This list was initiated by Agn$\grave{e}$s Acker and Pascal Le D\^{u} and regularly updated and published on the journal "L'astronomie" from the French Astronomers Society (SAF). It is also available with the VizieR2 tool\footnote{VizieR Service for Astronomical Catalogues: \url{https://vizier.u-strasbg.fr/viz-bin/VizieR-2}} from Strasbourg astronomical Data Center. Whenever the characteristic emission lines in a planetary nebula candidate's spectrum can be experimentally verified, the candidate's data is transmitted to the Hong Kong University to be included into the HASH PN database\footnote{The University of Hong Kong/Australian Astronomical Observatory/Strasbourg Observatory H-alpha Planetary Nebula (HASH PN) database: \url{http://202.189.117.101:8999/gpne/index.php}}.

\section{Instrumentation}
The campaigns were carried out on the C2PU telescopes of Calern site\footnote{C2PU 1m telescopes from l'observatoire de la  C\^{o}te d'Azur: \url{http://https://c2pu.oca.eu}}with a PNST\footnote{Planetary Nebulae Spectro Trackers: \url{http://planetarynebulae.net}} team. The telescope named Omicron is used during the 2017 and 2019 campaigns. In 2018, the telescope Epsilon, his twin, is used.
Both telescopes have a 1m-aperture, with f/d ratios reduced to 7 by using focal reducers. The spectrograph used for all campaigns is a LISA\footnote{Spectrograph LISA from Shelyak Instrument: \url{http://www.shelyak.com}} equipped  with a 50$\mu$m slit and a CCD ATIK414EX\footnote{ATIK CCD: \url{http://www.atik-cameras.com}} with pixels size of 6.45$\mu$m in binning 1x1. The final spectroscopic resolution we achieved with this instrument is about R=500.
Spectrum processing and reduction is performed with ISIS\footnote{ISIS software available here: \url{http://www.astrosurf.com/buil/isis-software.html}} software. The wavelength calibration was performed with an Argon/Neon lamp with a resulting calibration error of about 0.1\AA. For each target, a correction of the instrumental response was applied with a standard star at the same altitude as the observed target.  

\section{Results}
During the campaigns, more than 50 objects were observed and four of them show redshifted spectral emission lines. The redshift z of each emission line can be calculated according to the non-relativistic equation:

\begin{equation}\label{eq:norelz}
z=\frac{(\lambda_{measured} - \lambda_0)}{\lambda_0}
\end{equation}

We use $\lambda_{measured}$ for the wavelength of the specific emission line in spectrum of the object, and $\lambda_0$ for the appendant wavelength at rest.

\subsection{Object App~1}
The first object we will investigate is named App~1 after its discoverer, Florent Appert. He has found the object in the planetary nebula candidate search on a sky-map.org image. The coordinates of App~1, taken from the sky-map.org picture, are RA: 22h 49m 20.23s and DEC: +46\degr 07\arcmin 37.17\arcsec. Using the known magnitudes of nearby stars, we extrapolate the magnitude estimate for App~1 between 19mag to 20mag. The App~1 spectrum is shown in Fig.1 in the wavelength range between 4000\AA~ and 7500\AA. We have used the same range for all spectra to allow an easy comparison. The spectrum of App~1 is quite noisy, the signal-to-noise ratios of the emission lines are in the range of 2 to 20 (Tab.1).
\newline

\begin{figure}
   \label{figure:1}
     \includegraphics[width=\columnwidth]{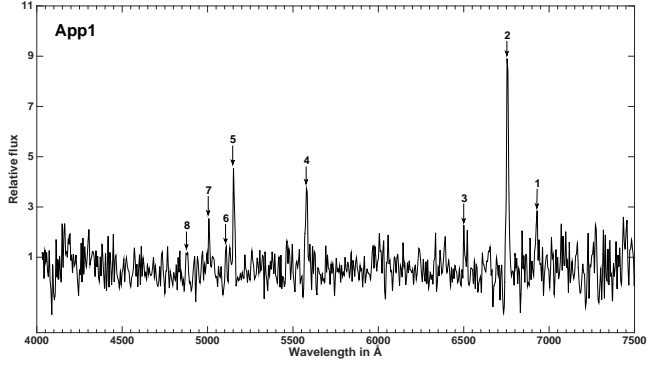}
     \caption{Spectrum of App~1 (date: 22nd September 2017, exposure time: 4 x 900sec)}
   \end{figure}

\begin{table}
\caption{App~1 results of the emission line measurements (enumeration refers to Fig.1).} 
\label{table:1} 
\begin{tabular}{l c c c c c} 
\hline
Line & $\lambda_{0}$ & $\lambda_{measured}$ & $\Delta\lambda$ & Flux & z \\  
  & \text{\AA} & \text{\AA} & \text{\AA} &   &   \\ 
\hline            
{1. [SII]} & 6731.3 &  6925.8 & 195.1 & 2.7 & 0.02889 \\
{2. H$\alpha$} & 6562.8 & 6756.2 & 193.3 & 8.8 & 0.02946 \\
{3. [SIII}] & 6312.1 & 6497.4 & 185.3 & 2.2 & 0.02935 \\
{4. HeII} & 5411.5 & 5576.5 & 165.0 & 3.7 & 0.03048 \\
{5. [OIII]} & 5006.9 & 5153.1 & 146.2 & 4.5 & 0.02920 \\
{6. [OIII]} & 4958.9 & 5106.6 & 147.7 & 1.3 & 0.02978 \\ 
{7. H$\beta$} & 4861.3 & 5007.3 & 146.0 & 2.5 & 0.03003 \\
{8. [ArIV]} & 4740.2 & 4877.9 & 137.7 & 1.1 & 0.02905 \\
\hline               
\end{tabular}
\end{table}

\subsection{Object Pre~21}
The second object in our planetary candidate list is named Pre~21 after Trygve Prestgard, who has discovered it using SDSS and WISE images. The coordinates of Pre~21 can be derived to RA: 18h 04m 19.62s and DEC: +00\degr 08\arcmin 04.96\arcsec. For Pre~21, we found a matching Gaia DR2 identification (Gaia DR2 source ID 4275388365469550720, with coordinates RA: 271.08166272300 and DEC: 0.13458491910). The mean magnitude in the Gaia DR2 passbands\footnote{Gaia photometric passbands: \url{https://www.cosmos.esa.int/web/gaia/iow_20180316}} for this source are (rp) 17.1mag, (bp) 17.5mag and (g) 18.7mag. This Gaia DR2 magnitude identifies Pre~21 as the brightest out of the four sources, which is also approved visually by comparing the PanSTARRS images.

 \begin{figure}
   \label{figure:2}
     \centering
     \includegraphics[width=\columnwidth]{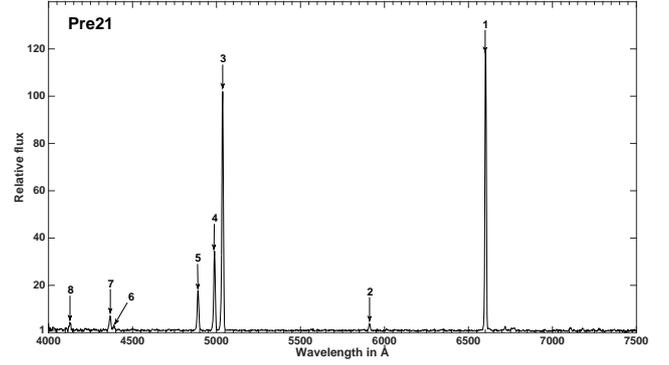}
     \caption{Spectrum of Pre 21 (date: 29th May 2019, exposure time: 2 x 1200sec)}
   \end{figure}

\begin{table}
\caption{Pre~21 results of the emission line measurements (enumeration refers to Fig.2).} 
\label{table:2}  
\begin{tabular}{l c c c c c}   
\hline
Line & $\lambda_{0}$ & $\lambda_{measured}$ & $\Delta\lambda$ & Flux & z \\  
  & \text{\AA} & \text{\AA} & \text{\AA} &   &   \\ 
\hline           
{1. H$\alpha$} & 6562.8 & 6601.9 & 39.05 & 118.8 & 0.00595 \\
{2. HeI} & 5876.0 & 5911,1 & 35.08 & 3.8 & 0.00597 \\
{3. [OIII]} & 5006.9 & 5036.1 & 29.30 & 100.1 & 0.00585 \\
{4. [OIII]} & 4958.9 & 4987.7 & 28.77 & 34.5 & 0.00580 \\
{5. H$\beta$} & 4861.3 & 4889.6 & 28.22 & 17.7 & 0.00580 \\
{6. [OIII]} & 4363.0 & 4389.3 & 26.30 & 3.1 & 0.00603 \\
{7. H$\gamma$} & 4340.5 & 4367.0 & 26.53 & 7.0 & 0.00611 \\
{8. H$\delta$} & 4101.7 & 4127.4 & 25.65 & 4.3 & 0.00625 \\
\hline     
\end{tabular}
\end{table}

\subsection{Object Pre~24}
The second object in our planetary candidate list is named Pre~24 after Trygve Prestgard, who has discovered it, using SDSS and WISE images. The coordinates of Pre~24 can be derived to RA: 04h 25m 53.63s and DEC: +39\degr 49\arcmin 19.69\arcsec. Using the magnitudes of nearby stars, we estimate the photometric magnitude for Pre~24 to lie in the range of 20mag to 21mag. 

 \begin{figure}
   \label{figure:3}
     \centering
     \includegraphics[width=\columnwidth]{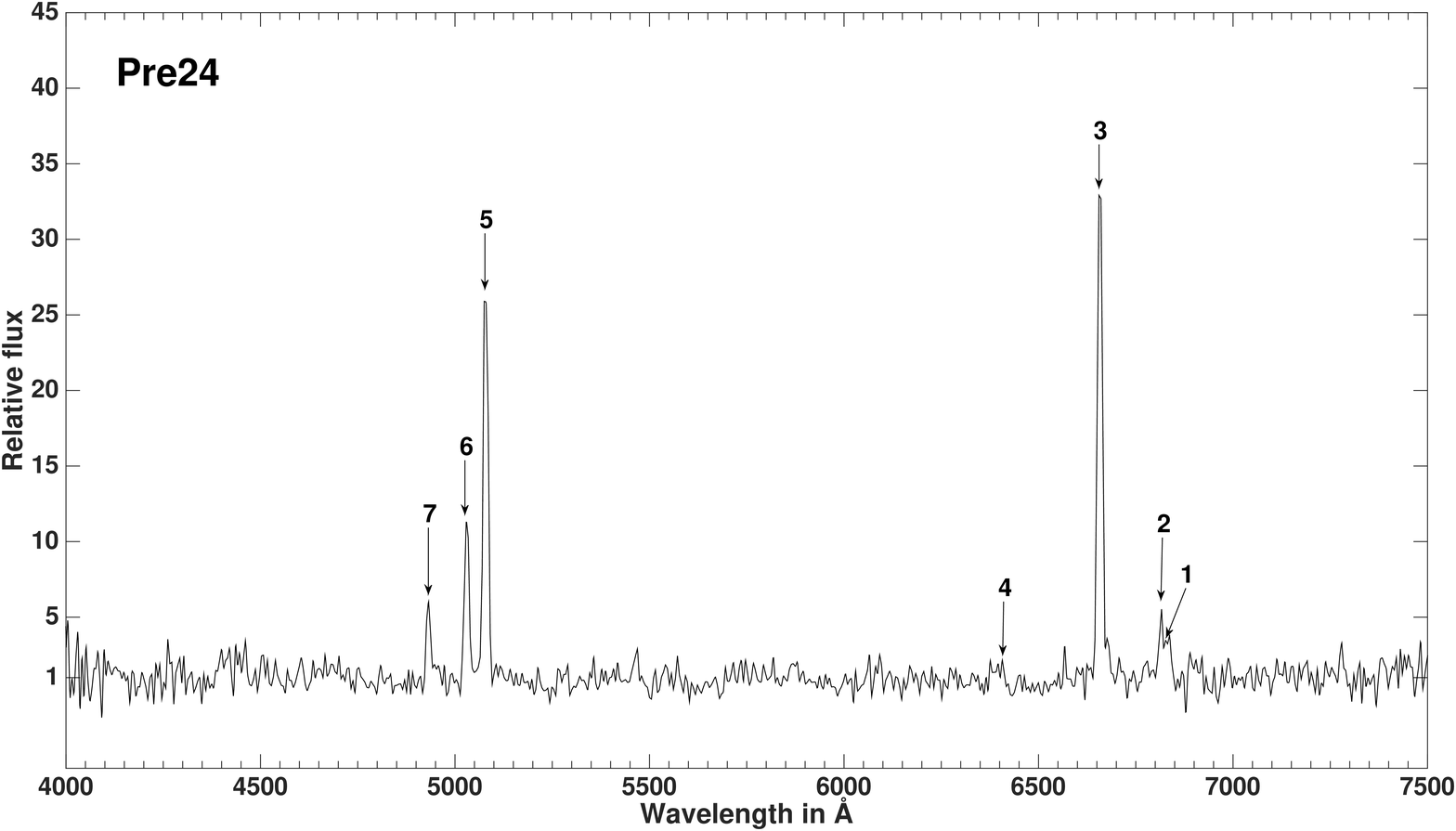}
     \caption{Spectrum of Pre 24 (date: 21st September 2017, exposure time: 4 x 900sec)}
   \end{figure}

\begin{table}
\caption{Pre~24 results of the emission line measurements (enumeration refers to Fig.3).} 
\label{table:3} 
\centering \begin{tabular}{l c c c c c}  
\hline
Line & $\lambda_{0}$ & $\lambda_{measured}$ & $\Delta\lambda$ & Flux & z \\  
  & \text{\AA} & \text{\AA} & \text{\AA} &   &   \\ 
\hline   
{1. [SII]} & 6731.3 & 6826.6 & 95.32 & 3.5 & 0.01416 \\
{2. [SII]} & 6717.0 & 6816.3 & 99.32 & 5.5 & 0.01478 \\
{3. H$\alpha$} & 6562.8 & 6658.0 & 95.18 & 32.8 & 0.01450 \\
{4. [SIII]} & 6312.1 & 6407.1 & 94.98 & 2.2 & 0.01505 \\
{5. [OIII]} & 5006.9 & 5078.9 & 72.07 & 25.9 & 0.01439 \\
{6. [OIII]} & 4958.9 & 5030.4 & 71.51 & 11.3 & 0.01442 \\
{7. H$\beta$} & 4861.3 & 4931.9 & 70.58 & 5.9 & 0.01452 \\
\hline               %inserts single line
\end{tabular}
\end{table}

\subsection{Object Ra~69}
The third object, we present, is named Ra~69. This planetary candidate was discovered by Thierry Raffaelli. He used SDSS and WISE images for this work. The coordinates of Ra~69 are RA: 19h 30m 23.64s and DEC: +37\degr 37\arcmin 06.58\arcsec. We have found a Gaia DR2 identification for Ra~69 with Gaia DR2 source ID 2051827137119772032, and with coordinates RA: 292.59844571250 and DEC: 37.61843556192. The mean magnitude in the Gaia DR2 passbands for Ra~69 are (rp) 17.8mag, (bp) 18.1mag and (g) 18.9mag.

\begin{figure}
     \label{figure:4}
     \centering
     \includegraphics[width=\columnwidth]{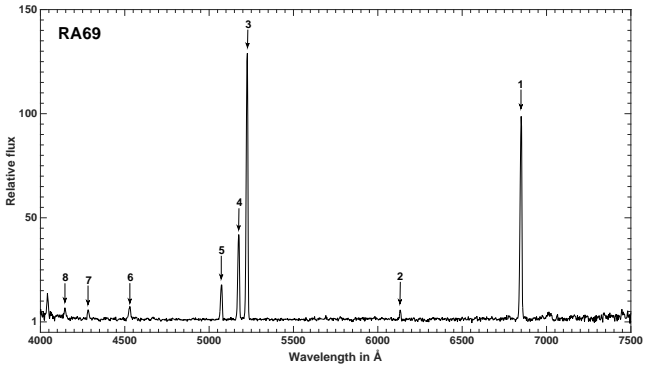}
     \caption{Spectrum of Ra~69 (date: 17th September 2017, exposure time: 4 x 900sec)}
\end{figure}

\begin{table}
\caption{Ra~69 emission line measurements (enumeration refers to Fig.4).} 
\label{table:4}  
\begin{tabular}{l c c c c c} 
\hline
Line & $\lambda_{0}$ & $\lambda_{measured}$ & $\Delta\lambda$ & Flux & z \\  
  & \text{\AA} & \text{\AA} & \text{\AA} &   &   \\ 
\hline              
{1. H$\alpha$} & 6562.8 & 6848.7 & 285.89 & 98.7 & 0.04356 \\
{2. HeI} & 5876.0 & 6132.7 & 256.67 & 5.7 & 0.04368 \\
{3. [OIII]} & 5006.9 & 5224.9 & 218.07 & 128.7 & 0.04355 \\
{4. [OIII]} & 4958.9 & 5174.9 & 215.94 & 41.9 & 0.04261 \\
{5. H$\beta$} & 4861.3 & 5072.7 & 211.32 & 17.8 & 0.04347 \\
{6. H$\gamma$} & 4340.5 & 4529.1 & 188.59 & 7.2 & 0.04345 \\
{7. H$\delta$} & 4101.7 & 4283.4 & 181.63 & 5.8 & 0.04428 \\
{8. H$\epsilon$} & 3970.1 & 4145.4 & 175.34 & 6.7 & 0.04417 \\
\hline                
\end{tabular}
\end{table}

\section{Discussion}
In this chapter, we will discuss the spectroscopic results in combination with the information in the survey images of the four objects. For each object, we will investigate the PanSTARRS survey picture to use its optical morphology to discuss extragalactic nature. In the centre of the PanSTARRS pictures, a purple cross defines the object coordinates and also indicates the position error caused by the slit width of the spectrograph.

In Tab.1, the emission lines measurements the object App~1 are shown, and for each emission line, the redshift is calculated using equation \eqref{eq:norelz}. For App~1, the redshift is z = 0.02953 $\pm$ 0.00053. From this redshift, we calculate a comoving radial distance of 126.4 Mpc. This and all other redshifts in this article are determined with the CosmoCal-online tool\footnote{Ned Wrights CosmoCal tool: \url{http://www.astro.ucla.edu/~wright/intro.html}} with the parameters $H_0$=69.6, $\Omega_M$=0.286 and $\Omega_{vac}$=0.714 (Wright 2006).

The emission line signature of App~1 indicates that this object could be a Seyfert galaxy (Osterbrok et al. 1976). The ratio of the [OIII]$\lambda$5007-flux and H$\beta$-flux is 1.8 (see Tab.1). Generally, the [OIII]$\lambda$5007/H$\beta$-ratio can be used as an indicator to differ a type 1 from a type 2 Seyfert galaxy (Dessauges-Zavadsky et al. 2000). The type 1 Seyfert galaxies have small ratios, typically with values [OIII]$\lambda$5007/H$\beta$ smaller than 3. 

Using modern SDSS classification data, the Seyfert assumption needs to be relativised. The emission-line ratios of [OIII]/H$\beta$, [NII]/H$\alpha$ and [SII]/H$\alpha$ can be used to distinguish between a Seyfert, a LINER or a HII galaxy. The log-ratios are the coordinates in diagrams which are commonly known as Baldwin-Phillips-Terlevich (BPT) diagrams (Kewley et al. 2006). We derive for the App~1 emission-line ratios the values log([OIII]$\lambda$5007/H$\beta$)=0.26 and log([SII]/H$\alpha$)=-0.51, identifying App~1 as a HII-region-like galaxy, and not a Seyfert galaxy, in the BPT-diagram. Particularly, the weak [SII] emission lines turn the classification and favours the HII nature of this galaxy. 

The PanSTARRs picture of App~1 (see Fig.5) unveils a diffuse core with a blueish loop that we interpret as a spiral arm structure of a galaxy. The diameter of the structure is about 0.7\arcmin~ which equals a projected diameter of 25 kpc, and this fits the typical size for a normal Spiral galaxy. Both, the spectroscopic and visual results are in accordance with the assumption that App~1 is a HII galaxy.

\begin{figure}
   \label{figure:5}
     \centering
     \includegraphics[width=\columnwidth]{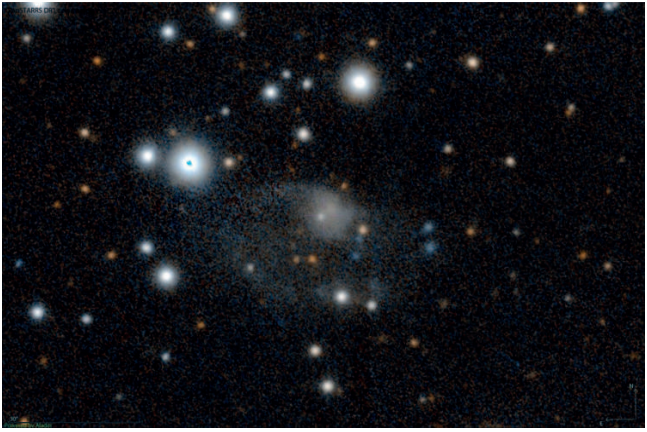}
     \caption{PanSTARRS DR1 image of App~1 (size: 2.4'x1.6', North up/East left).}
   \end{figure}
   
The results of the second object, Pre~21, are shown in Tab.2, with a mean value of the redshift of z = 0.00597 $\pm$ 0.00016. We conclude a comoving radial distance of 25.6 Mpc. With the visual object diameter the projected source size is about 2100 lyrs.

Also, the PanSTARRS image of Pre~21 shows a streak of light pointing to the north, with a projected length of about 20 klyrs. We would interpret this feature as a possible tidal stream, and Pre~21 as a small dwarf galaxy that could have survived a close encounter with another galaxy.

The emission-line ratio is log([OIII]$\lambda$5007/H$\beta$)=0.26. As the [SII] emission-lines are weak, the 2nd classification ratio can only be roughly estimated to log([SII]/H$\alpha$)<-1, and this attests Pre~21 to be probably a HII-region-like galaxy.

\begin{figure}
   \label{figure:6}
     \centering
     \includegraphics[width=\columnwidth]{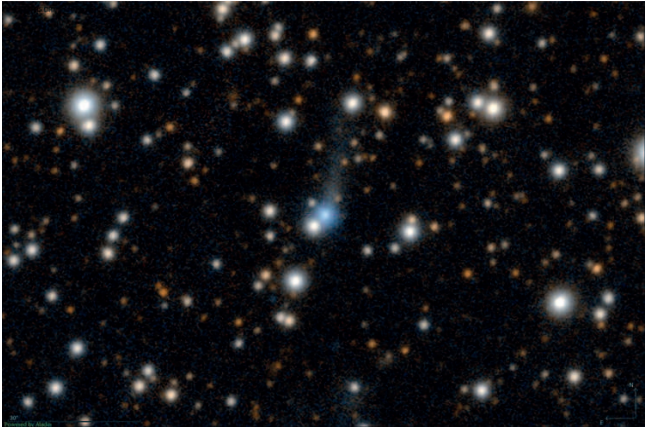}
     \caption{PanSTARRS DR1 image of Pre~21 (size: 2.4'x1.6', North up/East left).}
   \end{figure}

The results of the object Pre~24 in Tab.3 yield a mean redshift of z = 0.01454 $\pm$ 0.00029, with a comoving radial distance of 62.4 Mpc. The ratio values are log([OIII]$\lambda$5007/H$\beta$)=0.64 and log([SII]/H$\alpha$)=0.89 and classify Pre~24 as a HII-region-like galaxy.

The PanSTARRS picture of Pre~24 (Fig.7) shows a diffuse object, elongated in the RA-direction with a projected diameter of about 13 kpc. This is significantly smaller than the standard spiral galaxy size, and we assume that this extragalactic object might have been a spiral galaxy, but it has undergone interactions with other galaxies, causing an irregular morphology. The result could be a active galaxy core, with a lost spiral structure, or the spiral arms are too weak to be detected in the image. 

\begin{figure}
   \label{figure:7}
     \centering
     \includegraphics[width=\columnwidth]{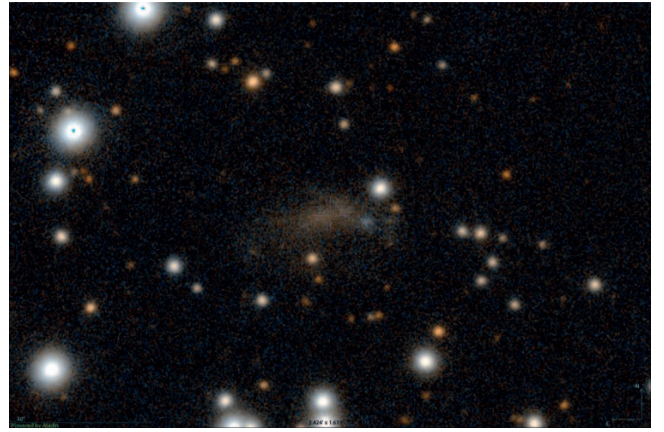}
     \caption{PanSTARRS DR1 image of Pre~24 (size: 2.4'x1.6', North up/East left).}
   \end{figure}
   
With the results of the fourth object Ra~69 (Tab.4), we derive a mean redshift of z = 0.04360 $\pm$ 0.00051. This redshift equals a comoving radial distance of 186.0 Mpc and the object visual diameter gives a projected diameter of about 13.000 lyrs. For Ra~69 no surrounding galactic structure can be seen in the PanSTARRS image (Fig.8). It's blueish and compact appearance is very similar to Pre~21, although Ra~69 is 4.7-times bigger than is "visual twin". This bright blue part could be seen as the bright bulge of a HII galaxy or Ra~69 could be a dwarf galaxy with a HII-region-like galactic core. 
The first emission-line ratio is log([OIII]$\lambda$5007/H$\beta$)=0.86 and the second ratio is estimated using the weak [SII]-signature with log([SII]/H$\alpha$)<-1.

\begin{figure}
   \label{figure:8}
     \centering
     \includegraphics[width=\columnwidth]{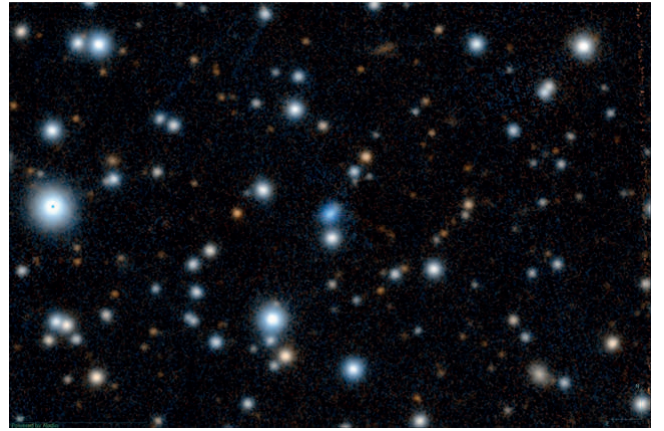}
     \caption{PanSTARRS DR1 image of Ra~69 (size: 2.4'x1.6', North up/East left).}
   \end{figure}

\section{Conclusion}
In the three spectroscopic campaigns we have done at the C\^{o}te d'Azur observatory at Calern in France, we have investigated the optical spectra of planetary nebula candidates. We have found extragalactic redshifts for four candidates and we have deduced their cosmological distances.  The candidates notated App~1 (RA: 22h 49m 20.23s, DEC: +46\degr 07\arcmin 37.17\arcsec) lies in a distance of 126 Mpc, the second candidate Pre~21 (RA: 18h 04m 19.62s, DEC: +00\degr 08\arcmin 04.96\arcsec) is 26 Mpc apart, the candidate named Pre~24 (RA: 04h 25m 53.63s, DEC: +39\degr 49\arcmin 19.69\arcsec) is in a distance of 62 Mpc and the fourth candidate is designed Ra~69 (RA: 19h 30m 23.64s, DEC: +37\degr 37\arcmin 06.58\arcsec) and its distance is 186 Mpc. 

For all four objects, the distances, which have been calculated from the measured redshifts of the emission lines, are not compatible with the possible range of distances of planetary nebula in the Milkyway, but with an extragalactic object nature. For all four objects, no broad emission lines are found, and the emission-line ratios of [OIII]$\lambda$5007/H$\beta$ and [SII]/H$\alpha$ allow a classification of the galaxy type using the BPT-diagrams and attest that App~1, Pre~21, Pre~24 and Ra~69 are highly probable HII-region-like galaxies. 
This conclusion is also in accordance with the result that we haven't found any radio catalogue records, which complies with the radio-quiet nature of HII galaxies.

We are fully aware of the fact that our equipment's capabilities are limited, for example by the low spectral dispersion of our spectra, so we cannot determine properly all emission-line ratios that are typically used for the BPT-classification method. But our intention was to show that amateur astronomers can contribute a small puzzle piece to extragalactic research, by persevering database research and own observation campaigns.

Also, it would be interesting to check whether deep optical images can be used to detect details of the galactic discs of the galaxies. Conclusively, we suggest further investigation of the four HII galaxies with professional telescopes.

\section*{Acknowledgements}
We would like to particularly thank Jean-Pierre Rivet from the C\^{o}te d'Azur observatory for allowing us to use these two C2PU telescopes at the Calern site.

\newpage

%%%%%%%%%%%%%%%%%%%%%%%%%%%%%%%%%%%%%%%%%%%%%%%%%%

% Don't change these lines
\bsp	% typesetting comment
\label{lastpage}
\end{document}